\documentclass{PoS}

\title{Investigating some technical improvements to glueball calculations.}

\ShortTitle{Glueball calculations}

\author{Biagio  Lucini\\
  College of Science, Swansea University,
  Singleton Park, Swansea SA2 8PP, UK\\
        E-mail: \email{b.lucini@swansea.ac.uk}}

\author{\speaker{Craig McNeile}\\\
Centre for Mathematical Sciences, Plymouth University, 
Plymouth, PL4 8AA, UK 
        Plymouth University\\
        E-mail: \email{craig.mcneile@plymouth.ac.uk}}

\author{Antonio Rago\\
Centre for Mathematical Sciences, Plymouth University, 
Plymouth, PL4 8AA, UK 
        Plymouth University\\
        E-mail: \email{antonio.rago@plymouth.ac.uk}}

\abstract{
We briefly discuss some issues concerning including the
pseudoscalar glueball interpolating operators into the
variational basis for computing the masses of the 
$\eta$ and $\eta^\prime$ mesons. As a start we present
some preliminary results for the correlators from pseudoscalar glueball
operators on $n_f$ = 2+1+1 twisted mass gauge configurations.
Preliminary results for the effect of open boundary conditions on the
masses of three glueballs computed in quenched QCD are presented. 
The statistics of glueball correlators are briefly discussed.
}

\FullConference{The 33rd International Symposium on Lattice Field Theory\\
		14 -18 July 2015\\
		Kobe International Conference Center, Kobe, Japan*}

\begin{document}

\section{Introduction}

The glueballs are truly non-perturbative objects in QCD (see~\cite{Mathieu:2008me} for a
review of glueballs). We
desperately hope they exist in nature, even if they are mixed in with
quark degrees of freedom, but, as many Beyond the Standard Model (BSM) experts are finding, experiment may ultimately disappoint us. The
quenched glueball spectrum was largely settled over a decade
ago~\cite{Morningstar:1999rf,Chen:2005mg}. 
Recent efforts have focused on unquenching the 
glueball spectrum~\cite{Gregory:2012hu}.
Even more recently,
it has been speculated, that the $0^{++}$ glueball in some 
strongly interacting BSM theories is
a candidate for a composite Higgs~\cite{Kuti:2014epa,Aoki:2014oha}.

In this paper we discuss some technical issues in lattice QCD studies of
glueball degrees of freedom, as part of our long term  goals  to find experimental evidence for
them and their possible contribution to BSM theories. In this paper we
stick with the SU(3) theory.

\section{Unquenching the pseuodoscalar glueball }

 There are a number of challenges with searching for glueball degrees 
of freedom in nature. Glueball operators will mix with quark
operators
with the same quantum numbers, so unquenched calculations
should include both glueball and quark disconnected diagrams in the
same variational calculation. Also high statistics are required.
Although large $N_c$ arguments suggest that the widths of glueballs
should be small, the candidate states which may include glueball
degrees of freedom usually have large widths. Hence, specialized
resonance techniques are required to study $0^{++}$ states, for example.

It may be easier to consider the pseudo-scalar glueball, even though
in the quenched theory the $0^{-+}$ glueball is the third heaviest
glueball. The widths of the $\eta$ and $\eta^\prime$ mesons are 
1.3 kev and 0.2 MeV
respectively, hence it is reasonable to neglect resonance effects.
Also there are fewer concerns about candidate molecules or tetraquarks
as there with the flavour singlet $0^{++}$ hadrons.

Some phenomenologically studies have suggested that in experiment
the pseudo-scalar glueball degrees of freedom 
are at a much lower mass than  the quenched 
pseudo-scalar glueball mass~\cite{Cheng:2008ss}. The few previous lattice studies
of unquenching the pseudo-scalar glueball have not seen much
difference in the mass at a fixed lattice
spacing~\cite{Gregory:2012hu,Richards:2010ck}. However, recently
the JLQCD collaboration have used the pseudo-scalar glueball
interpolating operators to exact
the mass of the $\eta^{\prime}$ meson~\cite{Fukaya:2015ara}, which 
suggests a strong mixing of the glueball with the $\eta^\prime$.

There has been a big effort to determine the mixing angle between the
$\eta$ and $\eta^\prime$ meson. For example, the LHCb experiment have
recently estimated $\eta$ and $\eta^\prime$ 
mixing~\cite{Aaij:2014jna}. Some $\eta - \eta^\prime$  mixing schemes also include the $0^{-+}$
glueball.  For example the KLOE experiment~\cite{Ambrosino:2006gk}
wrote the physical $\eta^\prime$ meson state in terms of 
light quarks ($\mid \overline{q} q \rangle$)
strange quarks ($ \mid \overline{s} s$)
and glueball degrees of freedom ($\mid glueball \rangle$).
\begin{equation}
\mid \eta^\prime \rangle = X_{\eta^\prime} \mid \overline{q} q \rangle 
                           + Y_{\eta^\prime} \mid \overline{s} s
                           \rangle 
+ Z_{\eta^\prime} \mid glueball \rangle 
\end{equation}
There are different parameterizations of the mixing angles (see~\cite{Cheng:2008ss}  for example), but KLOE used one with
the constraint   $X_{\eta^\prime}^2 + Y_{\eta^\prime}^2 = 1 $
The KLOE experiment fitted
$X_{\eta^\prime}$, $Y_{\eta^\prime}$, and $Z_{\eta^\prime}$
  to experimental branching fractions and obtained
$ Z_{\eta^\prime}^2 = 0.14 \pm 0.04 $.

There have been 
a number of lattice QCD calculations of $\eta^\prime - \eta$ 
mixing~\cite{McNeile:2000hf,Christ:2010dd,Gregory:2011sg,Dudek:2013yja,Michael:2013gka}.
So it would be good to include the glueball interpolating operators as
well, to try to determine $ Z_{\eta^\prime}$.
Although the variational basis used to study $\eta$ and $\eta^{\prime}$
mesons can be extended to include glueball interpolating operators.
At this moment, it is less clear to us how to define the quantity 
$ Z_{\eta^\prime}$ in terms of lattice QCD correlators.
Matrix elements of glueballs have been estimated using quenched QCD~\cite{Chen:2005mg}.

\begin{table}[tb]
\centering
\begin{tabular}{|c|c|c|c|c|c|c|c|} \hline
Ensemble & $\beta$ & $L^3 \times T$ & $\kappa$  & $\mu_l$  &
$\mu_{\sigma}$  & $\mu_{\delta}$  &   No. \\
\hline
D45.32sc  & 2.10 & $32^3 \times 64$ & 0.156315  & 0.0045 & 0.0937 & 0.1077 & 1100  \\
B55.32  & 1.95 & $32^3 \times 64$ & 0.161236  & 0.0055 & 0.135  & 0.17 & 2200  \\
A80.24s & 1.90 & $24^3 \times 48$ & 0.163204 & 0.008  & 0.15   &
0.197 & 2433  \\
\hline 
\end{tabular}
\caption{Parameters of twisted mass ensembles used in the glueball
  analysis.
Further details are in the papers~\cite{Baron:2010bv,Michael:2013vba} }
\label{tb:paramETMC}
\end{table}

We decided to use twisted mass configurations,
from the European Twisted Mass
collaboration,
with 
$n_f=2+1+1$ sea quarks, because these had successfully been used
to calculate the mass of the $\eta$ and $\eta^\prime$ 
mesons~\cite{Michael:2013gka},
so we were hopeful that the statistics would be high enough to
get a signal with glueball interpolating operators.
Three twisted mass ensembles were 
analyzed using the glueball code developed in~\cite{Lucini:2010nv}.
The parameters of the ensembles used are in table~\ref{tb:paramETMC}.

\begin{figure}
\centering
\includegraphics[%
  scale=0.3,
  angle=0,
  origin=c]{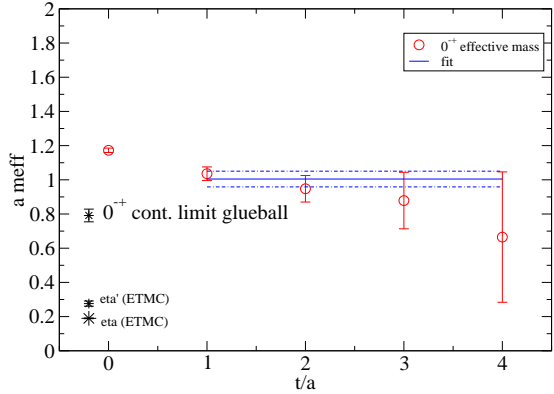}
\caption{Preliminary effective mass from the $0^{-+}$ glueball
  interpolating
operators from the B45.32sc ensemble. Also included is the continuum
limit glueball mass from quenched QCD~\cite{Chen:2005mg}. Also the
masses
of the $\eta$ and $\eta^\prime$ meson from 
ETMC~\cite{Michael:2013gka} from the same ensemble
are plotted.}
\label{fg:zeromp}
\end{figure}

Our glueball code includes an estimate of the projection of the state
onto the unphysical toleron state. Unfortunately, the B55.32 ensemble
had a very strong projection of the ground state in the pseudo-scalar 
channel to the toleron. The  D45.32sc  ensemble had a good overlap of the 
glueball operator with the ground state in the pseudo-scalar channel.
The physical box size of the D45.32sc ensemble was 2.0 fm, compared to 2.5
fm for the B55.32 ensemble. Normally, the contribution of the tolerons
is a finite volume effect, so it is not clear why the toleron
contribution
is larger for the ensemble with the larger physical size.
In figure~\ref{fg:zeromp} we plot our preliminary results for the 
effective mass from the $0^{-+}$ glueball operators from the B45.32sc ensemble.

\section{Open boundary conditions}

One concern about Monte Carlo calculations is that the autocorrelation
time is underestimated and the resulting final errors on physical
quantities
are underestimated, or even worse, there are systematic errors.
Different physical quantities have different autocorrelation times.
The topological charge is one quantity that has been observed to
have large autocorrelation effects which get worse as the continuum
limit is taken. Of particular concern are the results from the MILC
collaboration~\cite{Bazavov:2012xda}, where the time series for the topological charge
look worse  as the lattice spacing is reduced. 
However, as the  MILC collaboration
points out, the topological susceptibility
from the calculation agrees
with theoretical expectations. The two finest lattice spacings
simulated by the MILC collaboration are crucial for reducing the
errors on phenomenological quantities involving heavy quarks,
so it is important to study the issue.

Recently L\"uscher and Schaefer  have proposed the use of 
open boundary conditions~\cite{Luscher:2011kk,Luscher:2012av}
to improve the sampling of the topological charge.
McGlynn and Mawhinney have recently developed a simple
model
to explain the diffusion of topological charge and compared it to
their lattice data~\cite{McGlynn:2014bxa}.
Full QCD calculations with open boundary conditions have already 
started~\cite{Soldner:2015oea}.

What is not clear is whether the large autocorrelation on
the topological charge is important for spectral quantities,
such as the masses of particles. 
Chowdhury et al. have recently made a number of studies of 
the pseudo-scalar~\cite{Chowdhury:2014mra}
and scalar~\cite{Chowdhury:2014kfa} glueballs comparing periodic and open
boundary conditions. Within their large 
errors the glueball masses computed with periodic and open boundary conditions 
are broadly consistent.

The errors on the mass of the $0^{++}$ glueball by Chowdhury et al.~\cite{Chowdhury:2014mra,Chowdhury:2014kfa}
range from 6\% to 17\%. Given that many high precision lattice QCD
calculations quote
results at under 1\%, then it is desirable to have a more accurate calculation.
The calculation of Chowdhury et al. only use one glueball operator
(smeared with the Wilson flow) for each channel, so one improvement
would be the use of a variational calculation.

\begin{table}[tb]
\centering
\begin{tabular}{|c|c|c|} \hline
$\beta$ & $L^3 \times T$  &  No. \\
\hline
6.0625  & $16^4$   & 11779  \\
6.338   & $24^4$   &  4485  \\
\hline 
\end{tabular}
\caption{Parameters of quenched calculation using the Wilson gauge action}
\label{tb:param}
\end{table}

In table ~\ref{tb:param}, we report the parameters of the quenched QCD
calculations, used to compute the masses of the glueballs with open
boundary
conditions.  The Wilson gauge action was used.
The configurations were generated with 
a standard
update algorithm based on an admixture of Cabibbo-Marinari heathbath and
over-relaxed sweeps.
In this calculation
we use a variational calculation, with the code, developed and applied
for large $N_c$ calculations~\cite{Lucini:2010nv} 
and unquenched calculations~\cite{Gregory:2012hu}.
The use of open boundary conditions means that some 
of the time slices of correlators have unphysical contributions.
In our initial calculations we only used one or two time slices
per correlators, rather than average the correlator over all time slices
as is normally done for glueball calculation with periodic boundary 
conditions.

\begin{figure}[t]
\centering
\includegraphics[scale=0.5]{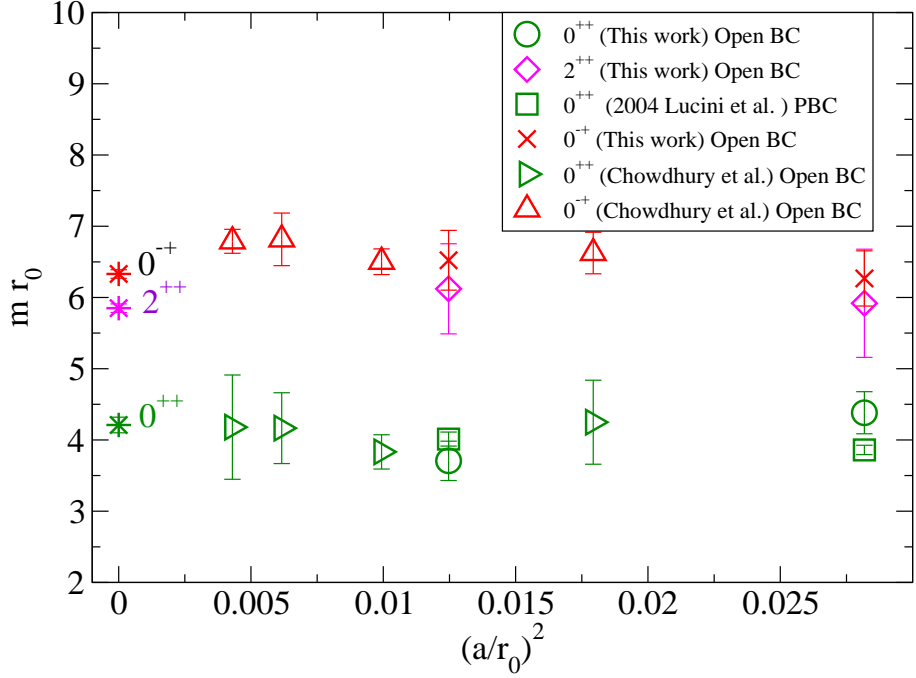}
\caption{Preliminary results for the masses of the $0^{++}$, $0^{-+}$
  and $2^{++}$ glueballs in units of $r_0$ computed with open
  boundary conditions, versus the square of the lattice spacing
in units of $r_0$.
We also include results for the glueball masses
  from Chowdhury et 
al.~\cite{Chowdhury:2014kfa,Chowdhury:2014mra} with open boundary
conditions. For comparison we include the glueball masses computed
with periodic boundary conditions (PBC) by Lucini  et
al. ~\cite{Lucini:2004my} and the continuum limit results from
Morningstar and Peardon~\cite{Morningstar:1999rf}.}
\label{fg:openRes}
\end{figure}

We use Sommer's $r_0$  parameter to determine the
the lattice spacing~\cite{Sommer:1993ce}.
Our preliminary results in figure~\ref{fg:openRes} 
seem to show the the results for glueball masses from open and periodic 
boundary conditions are consistent, but the errors still need to be
reduced for a definite high precision conclusion.
The use of the
variational code has not reduced the statistical errors over
those of Chowdhury et al.~\cite{Chowdhury:2014kfa}. We have probably
been too conservative in our choice of time slices to average over.
We are currently investigating other options.

\section{Statistical distribution of glueball correlators}

Calculations which include glueball degrees of freedom
      require large statistics. One concern is that the
probability distribution of the correlators has a ``long tail.''
This is a potentially generic feature of disconnected diagrams.
See for example the modeling of the correlators of the $\eta^{\prime}$ 
meson in~\cite{Gregory:2011sg}. 

In some physical calculations, such as the moments in heavy ion
collisions, the probability distribution of the data is 
important~\cite{Chen:2015vra}. In standard lattice QCD calculations,
the law of large numbers makes the underlying 
distribution of correlators irrelevant, if the
statistics are large enough to be in the asymptotic regime.
One motivation for studying the statistical distribution of the
glueball correlators is the possible use
of the techniques developed
Endres et al.~\cite{Endres:2011mm} for many body simulations,  where a ``filter like procedure''
is applied to dramatically reduce the statistical errors.

\begin{figure}[t]
\centering
\includegraphics[scale=0.5]{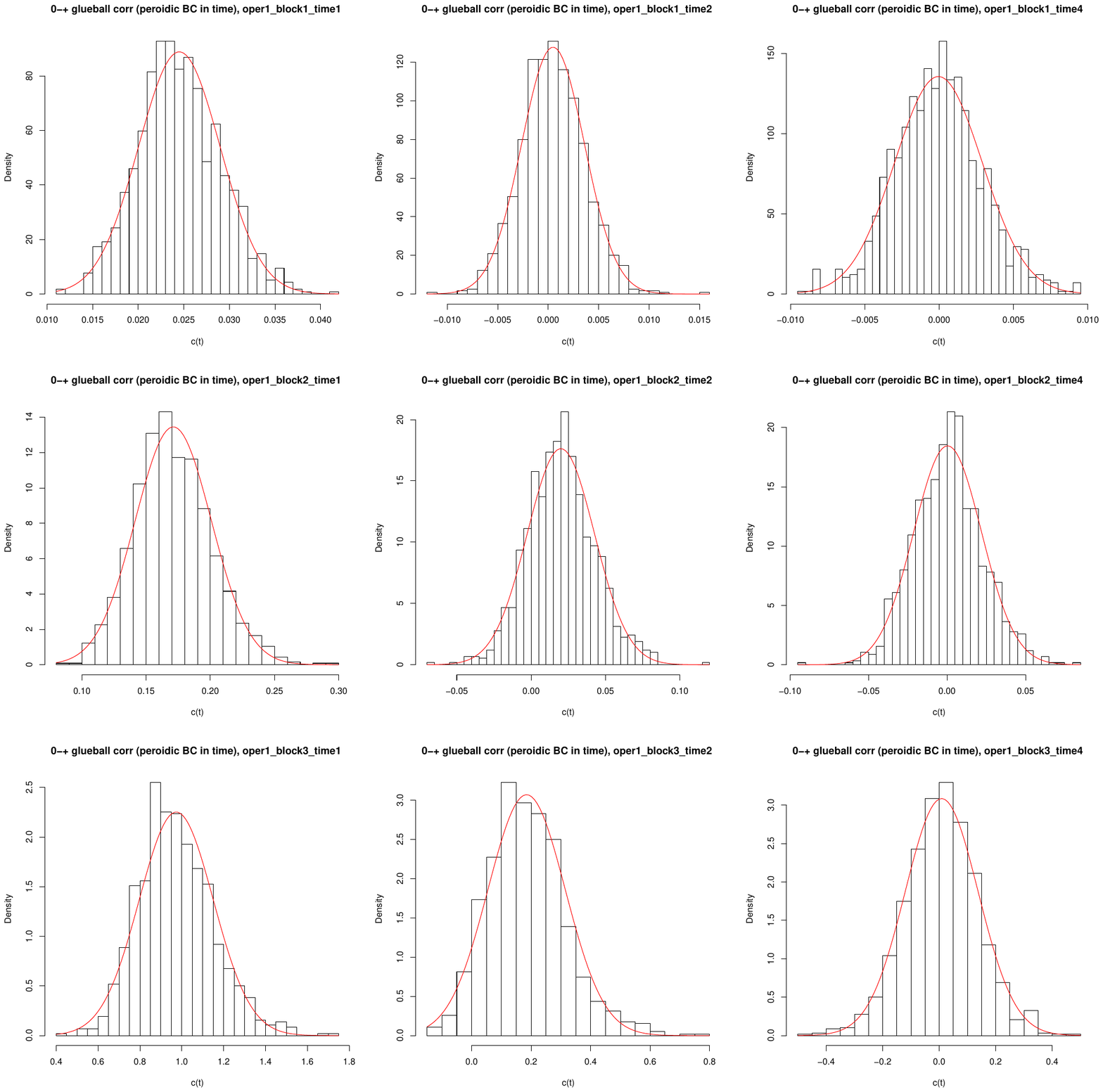}
\caption{Some $0^{-+}$ glueball correlators, for a specific
  operator, from the B55.32 twisted mass
  ensemble. The
different graphs are for the time-slices (1 to 3) and 
blocking levels (1 to 3).}
\label{fg:corrDist}
\end{figure}

In figure~\ref{fg:corrDist} we plot the histograms of 
some  $0^{-+}$ glueball correlators from
the B55.32 twisted
mass
ensemble. Also included is a fit with  a Gaussian fit model made with the
the R statistical system~\cite{Rref}. The Gaussian distribution gives a good
description of the correlator.

\section{Conclusions}

We have started a project to study whether pseudo-scalar glueball
interpolating operators
strongly couple to the $\eta^\prime$ meson. 
We have presented preliminary results for the effects of open
boundary conditions on the three lightest glueballs. 

This work
is supported by STFC under the DiRAC framework. We are grateful for
the support from the
HPCC Plymouth, where part of the numerical computations have been carried out.
We thank GridPP and Scotgrid for their support.
The ETMC configurations were downloaded from 
the ILDG~\cite{Beckett:2009cb}. Biagio  Lucini is supported by STFC (grant ST/L000369/1)
Antonio Rago is supported by the Leverhulme Trust (grant RPG-2014-118) and STFC
(grant ST/L000350/1).

%%\bibliographystyle{h-physrev5}
%%\bibliography{open} 

\end{document}